\documentstyle[12pt]{article}
\begin{document}
\begin{center}
{\noindent{\bf{Matsubara Frequency Sums }}}

\vspace{0.5cm}

Alok Kumar{\footnote{e-mail address: alok@iiserbhopal.ac.in}} \\
IISER, Bhopal\\
ITI Campus (Gas Rahat) Building\\
Govindpura, Bhopal - 23 \\
India. \\
\end{center}

\vspace{3.5cm}            

{\noindent{\it{Abstract}}}

We cannot use directly the results of zero-temperature at finite temperature, for at  finite temperature the average is to be carried over all highly degenerate excited states unlike zero-temperature average is only on unique ground state. One of the formal way to take into account the finite temperature into quantum field theory is due to Matsubara, to replace temporal component of eigenvalues $k_{4}$  by $\omega_{n}=\frac{2\pi n}{\beta}$ $\left(\frac{2\pi (n+\frac{1}{2})}{\beta}\right)$ with summation over all integer values of $n$. The summation is done with the infinite series expansion of $\coth (\pi y)$. With the chemical potential $\mu$, $\omega_{n}$ will be replaced by  $\omega_{n} - \mu$ in the eigenvalues and the summation over $n$ cannot be done easily. Various methods exist to evaluate it. We use the infinite series expansion of $\coth (\pi y)$ to work operationally  for such Matsubara frequency sums. 
 
\vspace{0.5cm}

\newpage 

Euclidean field theory at zero-temperature is extended to  the finite temperature with  Matsubara's imaginary time formalism [1]. At zero-temperature, it is convenient to have analytical continuation from real time to imaginary time : $t\rightarrow i\tau$ or $x^{0}\rightarrow i x^{4}$ to make  Minkowski space-time, Euclidean. Euclidean partition function $Z$ in the path integral formalism 
\begin{equation}
Z=\int [dA]e^{\int d^4 x \mathcal{L}}
\end{equation}
is performed to get effective action. Effective action of the form $Tr\log det (operator)$ under Gaussian approximation [2]. Under the spell of Matsubara formalism the $k_{4}$ is replaced by $\frac{2\pi n}{\beta}$ for bosons and by $\frac{2\pi (n + \frac{1}{2})}{\beta}$ for fermions and the integration over  $k_{4}$ is replaced by the sum over all integer values of $n$. We consider bosons and introduce  $\omega_{n}=\omega_{0}n$, where $\omega_{0}= \frac{2\pi }{\beta}$. Savvidy vacuum of QCD at finite temperature has been studied by several authors [2-9] with Matsubara frequency sums. To evaluate typical Matsubara frequency sums [10]
\begin{equation}
S=\displaystyle\sum_{n= -\infty}^{\infty}\ln(\omega_{n}^2 + \omega^2)
\end{equation}
we use the trick
\begin{equation}
\frac{d S}{d \omega}=\displaystyle\sum_{n= -\infty}^{\infty}\frac{2\omega}{(\omega_{n}^2 + \omega^2)}= \displaystyle\sum_{n= -\infty}^{\infty}\left(\frac{1}{\omega_{0}}\right)\frac{2\left(\frac{\omega}{\omega_{0}}\right)}{(n^2 + (\frac{\omega}{\omega_{0}})^2)}.
\end{equation}
We use the infinite series expansion of $\coth(\pi y)$ [14-18]
\begin{equation}
\displaystyle\sum_{n= -\infty}^{\infty}\frac{y}{(n^2 + y^2)}=\pi \coth(\pi y)
\end{equation}
With the equation (4), $S$,
\begin{equation}
\left.
\begin{array}{ccc}
S&=&2\int d\left(\pi\frac{\omega}{\omega_{0}}\right) \coth(\pi \frac{\omega}{\omega_{0}}) \\
&=&2\int \frac{d\left({e^{\left(\pi\frac{\omega}{\omega_{0}}\right)}-e^{-\left(\pi\frac{\omega}{\omega_{0}}\right)}}\right)}{{e^{\left(\pi\frac{\omega}{\omega_{0}}\right)}-e^{-\left(\pi\frac{\omega}{\omega_{0}}\right)}}}\\
&=&2 \ln\left(e^{\left(\pi\frac{\omega}{\omega_{0}}\right)}-e^{-\left(\pi\frac{\omega}{\omega_{0}}\right)} \right).\\
\end{array}
\right\}
\end{equation}
Inclusion of chemical potential $\mu$, $\omega_{n}$ will be replaced by  $\omega_{n} - \mu$ in the eigenvalues [2-9], equation (3)
\begin{equation}
\frac{d S}{d \omega}=\displaystyle\sum_{n= -\infty}^{\infty}\frac{2\omega}{((\omega_{n}-\mu)^2 + \omega^2)}.
\end{equation} 
We cannot use equation (4) directly to evaluate the summation in equation (6).  This summation  are evaluated in [12] using contour integration and in [11, 13] using some particular trick. We use the infinite series expansion of  $\coth(\pi y)$  to evaluate this summation with modification and it works operationally for any complex chemical potential. We consider equation (6) with the complex chemical potential $z_0= \mu_{R} + i\mu_{I}$ 
\begin{equation}
\frac{d S}{d \omega}=\displaystyle\sum_{n= -\infty}^{\infty}\frac{2\omega}{((\omega_{n}-z_{0})^2 + \omega^2)}.
\end{equation}
We can always factorise the equation (7) to have the power of $n$ in the denominator 1,
 \begin{equation}
\displaystyle\sum_{n = -\infty}^{\infty}\frac{2\omega}{(\omega_{n}-z_{0})^2+\omega^2} = \displaystyle\sum_{n = -\infty}^{\infty}\frac{1}{\omega + i \omega_{n}- i z_{0} }+\displaystyle\sum_{n = -\infty}^{\infty}\frac{1}{\omega-i \omega_{n} + i z_{0} }.
\end{equation}
We rewrite the equation (8)
 \begin{equation}
\displaystyle\sum_{n=-\infty}^{\infty}\frac{2\omega}{(\omega_{n}-z_{0})^2+\omega^2} = \frac{1}{\omega_{0}}\left[\displaystyle\sum_{n=-\infty}^{\infty}\frac{1}{\left(\frac{\omega}{\omega_{0}}\right) - i \left(\frac{z_{0}}{\omega_{0}}\right) + i n}+\displaystyle\sum_{n=-\infty}^{\infty}\frac{1}{\left(\frac{\omega}{\omega_{0}}\right) + i \left(\frac{z_{0}}{\omega_{0}}\right)- i n }\right].
\end{equation}
We rewrite the expansion $\coth(\pi y)$ in equation (4) to have the power of $n$ in the denominator 1,
\begin{equation}
\left.
\begin{array}{ccc}
\displaystyle\sum_{n=-\infty}^{\infty}\frac{1}{y+i n} + \displaystyle\sum_{n=-\infty}^{\infty}\frac{1}{y-i n}&=& 2\pi\coth(\pi y);\\
\displaystyle\sum_{n=-\infty}^{\infty}\frac{1}{y+i n} &=& \displaystyle\sum_{n=-\infty}^{\infty}\frac{1}{y-i n}\\
&=& \pi\coth(\pi y).\\
\end{array}
\right\}
\end{equation}
With equation (9) and equation (10), the equation (7) reads
 \begin{equation}
\frac{d S}{d\left(\pi \frac{\omega}{\omega_{0}}\right)}= \left[\coth\pi\left(\left(\frac{\omega}{\omega_{0}}\right) - i \left(\frac{z_{0}}{\omega_{0}}\right)\right) +  \coth\pi\left(\left(\frac{\omega}{\omega_{0}}\right) + i \left(\frac{z_{0}}{\omega_{0}}\right)\right)\right]
\end{equation}
The solution of equation (11) is similar to equation (5)
\begin{equation}
\begin{array}{ccc}
S&=&\left[\ln\left(e^{\pi\left(\left(\frac{\omega}{\omega_{0}}\right) - i \left(\frac{z_{0}}{\omega_{0}}\right) \right)}-e^{-\pi\left(\left(\frac{\omega}{\omega_{0}}\right) - i \left(\frac{z_{0}}{\omega_{0}}\right) \right)}\right)+ \ln\left(e^{\pi\left(\left(\frac{\omega}{\omega_{0}}\right) + i \left(\frac{z_{0}}{\omega_{0}}\right) \right)}-e^{-\pi\left(\left(\frac{\omega}{\omega_{0}}\right) + i \left(\frac{z_{0}}{\omega_{0}}\right) \right)}\right) \right]\\
&=&\left[2\pi\left(\frac{\omega}{\omega_{0}}\right) + \ln\left(1-e^{-2\pi\left(\left(\frac{\omega}{\omega_{0}}\right) - i \left(\frac{z_{0}}{\omega_{0}}\right) \right)}\right)+ \ln\left(1 - e^{-2\pi\left(\left(\frac{\omega}{\omega_{0}}\right) + i \left(\frac{z_{0}}{\omega_{0}}\right) \right)}\right) \right].
\end{array}
\end{equation}
As a special case, we take $z_{0} = i \mu$, and  for $\omega_{0}= \frac{2\pi }{\beta}$, $S$ in equation (12)
\begin{equation}
S = \omega \beta + \ln\left(1 - e^{\beta\left( \omega -\mu\right)}\right) +\ln\left(1 - e^{\beta\left( \omega + \mu\right)}\right).
\end{equation}
The expression for  $S$ in equation (13) matches with the expression for  $\Omega$ in equation (3.63) of reference [12].

To summarise, we factorise the summation to be performed  and  $\coth(\pi y)$ expansion so that  the power of $n$ in the denominator is one and by doing this chemical potential separate out as constant term from $n$. This method works operationally for  the range of summation $n = -\infty$ to  $n = \infty$ or for the range of summation can be made $n = -\infty$ to  $n = \infty$. In the Appendix we give more identities closely related to  $\coth(\pi y)$ with the hope this method can extended to other kinds of Matsubara sums with only care about the convergence of the series. It is shown to be true for one known example. For more mathematical details regarding Matsubara frequency sums one can  refer  to [19]. 

\vspace{0.5cm}

{\noindent{\bf{Acknowledgements}}}

\vspace{0.5cm}

I thank Director, IISER, Bhopal  for providing constant help and encouragement during the completion of this work. Discussion with Prof. R. Parthasarathy (CMI, Chennai) is acknowledged with thanks. 
 
\newpage 

\vspace{0.5cm}

{\noindent{\bf{Appendix}}}

\vspace{0.5cm}

The infinite series expansion of $\coth(\pi y)$ is  a special case 
\begin{equation}
\displaystyle\sum_{n=-\infty}^{\infty}\frac{a^{2l-1}}{n^{2 l}+ a^{2 l}}=\frac{\pi}{2 l}\displaystyle\sum_{k = -l}^{l-1}e^{\left(\frac{(2k+1)}{2l}\right)\pi i}\cot\left(\pi ae^{\left(\frac{(2k+1)}{2l}\right)\pi i}\right)
\end{equation}
where $a$ is real and positive and $l$ is a positive integer [15]. For $l=1$ 
\begin{equation}
\displaystyle\sum_{n= -\infty}^{\infty}\frac{y}{(n^2 + y^2)}=\pi \coth(\pi y).
\end{equation}
The following identities follow from equation (10)
\begin{equation}
\left.
\begin{array}{ccc}
\displaystyle\sum_{n= -\infty}^{\infty}\frac{n}{(n^2 + y^2)}&=& 0\\
\displaystyle\sum_{n= -\infty}^{\infty}\frac{1}{(x-i n)(y -i n)}&=& \frac{\pi(\coth(\pi y)-\coth(\pi x))}{x-y}\\
\displaystyle\sum_{n= -\infty}^{\infty}\frac{1}{(x+i n)(y -i n)}&=& \frac{\pi(\coth(\pi y)+\coth(\pi x))}{x+y}\\
\end{array}
\right\}
\end{equation}
We differentiate both sides of equation (10) to get
\begin{equation}
\left.
\begin{array}{ccc}
\displaystyle\sum_{n= -\infty}^{\infty}\frac{1}{(y+ i n)^2}= \frac{\pi^2}{\sinh^2(\pi y)} \\
\displaystyle\sum_{n= -\infty}^{\infty}\frac{1}{(y- i n)^2}= \frac{\pi^2}{\sinh^2(\pi y)} .\\
\end{array}
\right\}
\end{equation}
With the equation (17) we can manipulate the following identities with the help of logarithmic function [14-18]
\begin{equation}
\displaystyle\sum_{n= -\infty}^{\infty}\frac{(-1)^{n}}{(y^2+n^2)}= \frac{\pi}{y\sinh(\pi y)} 
\end{equation}
 \begin{equation}
y\displaystyle\prod_{n= 1}^{\infty}\left(1+\frac{y^2}{n^2\pi^2}\right)= \sinh(\pi y)
\end{equation}
\begin{equation}
\displaystyle\prod_{n= -\infty}^{\infty}\left[1+\left(\frac{x}{2\pi n+y}\right)^2\right]= \frac{\cosh(\pi x)-\cos(y)}{1-\cos(y)} 
\end{equation}

\newpage

{\noindent{\bf{References}}}

\vspace{0.5cm}

\begin{enumerate}
\item T.\,Matsubara,\,Prog.\,Theor.\,Phys.,\,{\bf{14}},\,351\,(1955).
\item R.\,Parthasarathy,\, and Alok\, Kumar,\, Physical Review{\bf{ D75}}, 085007 (2007).
\item J.\,J.\,Kapusta, Nucl. Phys. {\bf{B190}}, 425 (1981);\\ 
      B. Muller and J. Rafelski, Phys. Lett. {\bf{B101}}, 111 (1981) .
\item J. Chakrabarti, Phys. Rev. {\bf{D24}}, 2232 (1981); \\
      M. Reuter and W. Dittrich, Phys. Lett. {\bf{B144}}, 99 (1984).
\item M.\,Ninomiya and N.\,Sakai,\,Nucl.\,Phys.\,{\bf{B190}},\,316\,(1981).
\item A.\,Cabo,\,O.\,K.\,Kalashnikov,\, and A.\,E.\,Shabad,\,Nucl.\,Phys.\,{\bf{B185}},\,473\,(1981).
\item A.\,O.\,Starinets,\,A.\,S.\,Vshivtsev,\,and V.\,C.\,Zhukovsky,\,Phys.\,Lett.\,{\bf{B322}},\,403\,(1994).
\item P.\,N.\,Meisiger and M.\,C.\,Ogilvie,\,Phys.\,Rev.\,{\bf{D66}},\,105006\,(2002).
\item M.\,Loewe,\,S.\,Mendizabal and J.\,C.\,Rojas,\,{\it{Background Field Method at Finite Temperature and Density,}}\,arXiv.hep-ph/0512042v1.
\item L.\,Dolan and R.\,Jackiw,\,Phys.\,Rev.\,{\bf{D9}},\,(1974).
\item J. J. Kapusta, {\it{Finite Temperature Field Theory}}, Cambridge
University Press, 1989.
\item Michel\, L.\, Bellac, {\it{Thermal Field Theory}}, Cambridge
University Press, 1986.
\item Ashok\, Das, {\it{Finite Temperature Field Theory}}, World Scientific, Singapore 1997.
\item  Arfken and  Hans J.\, Weber,\, {\it{Mathematical Methods for Physicists (sixth edition)}},\,Academic Press,\,(2005).
\item D.\,S.\, Mitrinovic and  J.\,H.\, Michael,\, {\it{Calculus of Residues}},\,P.\,Noordhoff\,Ltd.\,-\,Groningen,\,(1965).
\item I.\,S.\,Gradshteyn and I.\,M.\,Ryzhik, \,{\it{Table of Integrals, Series and
Products}},\, Academic Press, \,(1965).
\item Alan Jeffrey,\,{\it{Handbook of Mathematical Formulas and Integrals}},\,Elsevier Academic Press,\,Third Edition\,(2004).
\item Daniel Zwillinger,\,{\it{Standard Mathematical Tables and Formulae}},\,CRC Press,(1996).
\item Olivier Espinosa ,\,{\it{On the Evaluation of Matsubara Sums}},\,arXiv:0905.3366v1[math.CA].
\end{enumerate}
\end{document}